\documentclass[submission,copyright,creativecommons]{eptcs}
\providecommand{\event}{QPL 2011}

\usepackage{amsthm}
\usepackage{amsfonts}

\usepackage[small,nohug,heads=vee]{diagrams}
\diagramstyle[labelstyle=\scriptstyle]

\newtheorem*{ntheorem}{Theorem}

\providecommand{\urlalt}[2]{\href{#1}{#2}} 
\providecommand{\doi}[1]{doi:\urlalt{http://dx.doi.org/#1}{#1}}

\title{Bohrification of local nets}
\author{Joost Nuiten 
\institute{Universiteit Utrecht} 
\email{j.j.nuiten@students.uu.nl}}

\def\titlerunning{Bohrification of local nets}
\def\authorrunning{Joost Nuiten}

\begin{document}

\maketitle

\begin{abstract}
Recent results by Spitters et. al. suggest that quantum phase space can usefully be regarded as a ringed topos via a process called Bohrification. They show that quantum kinematics can then be interpreted as classical kinematics, internal to this ringed topos.

We extend these ideas from quantum mechanics to algebraic quantum field theory: from a net of observables we construct a presheaf of quantum phase spaces. We can then naturally express the causal locality of the net as a descent condition on the corresponding presheaf of ringed toposes: we show that the net of observables is local, precisely when the presheaf of ringed toposes satisfies descent by a local geometric morphism.
\end{abstract}

\section*{Introduction}
\addcontentsline{toc}{section}{Introduction}
A decade ago, it was suggested by Butterfield and Isham \cite{bi98} that topos theory could provide a better framework for formulating something like quantum logic. They noticed that while a quantum phase space does not exist as an ordinary topological space, it does exist as a topos with an internal `space', or locale. This perspective allowed them to give a geometric formulation of the Kochen-Specker theorem, which then precisely stated that this internal phase space had no global points.

Recently, this idea has been heavily extended by D\"oring and Isham, who go as far as trying to find a complete topos-theoretic foundation for physics in their sequence of articles \cite{id07a}-\cite{id07d}. In particular, in \cite{id07b} they show how to realize quantum kinematics internal to a topos, using a process called daseinisation and an internal locale called the spectral presheaf.

Inspired by this, Spitters et al. \cite{hls09} provided a similar, but different description of a quantum phase space as a topos with a locale, using a procedure they called Bohrification. Equivalently, one may view this phase space as a ringed topos, or even a topos with an internal C${}^*$-algebra. Such a description of a phase space fits in the modern approach to geometry, which traces back to Grothendieck who noticed that a geometric space could be formalized by a (locally) ringed topos. Presently, the most advanced theory of general geometry (as discussed in \cite{lu09}) all revolves around regarding locally ringed (higher) toposes as generalized spaces.

Where the spectral presheaf is used to describe quantum kinematics in \cite{id07b}, the authors of \cite{hls09} use Bohrification to describe quantum kinematics in a way that strongly resembles the description of classical kinematics. In this text we will extend the Bohrification process from quantum mechanics to quantum field theory: using the formalism of AQFT to describe a quantum field theory on a spacetime $X$
in terms of a copresheaf of algebras of observables, we construct a presheaf of ringed toposes on the opens of $X$, called a Bohrified net on the spacetime $X$.  

Our main result is then a characterization of the causal locality of quantum
field theory in these terms: we show that this Bohrified net of
toposes on spacetime satisfies -- over any spatial hyperslice --
descent by local geometric surjections precisely if it comes from a
causally local net.

\section*{Bohrification}
\addcontentsline{toc}{section}{Bohrification}
The main inspiration for the Bohrification process is the observation that a lot of information about a C${}^*$-algebra is contained in its commutative subalgebras. Indeed, in two major theorems about the structure of quantum mechanics, commutative subalgebras play a main role. The first theorem, by Kochen and Specker \cite{ks67}, states that for a Hilbert space $H$ of dimension greater than 2, there exists no map $B(H)\rightarrow \mathbb{C}$ for which the restriction to each of the commutative subalgebras of $B(H)$ is a ${}^*$-homomorphism. Furthermore, a famous theorem by Gleason \cite{gl57} characterizes the states on $B(H)$ by their restrictions to its commutative subalgebras: if $\rho: B(H)\rightarrow \mathbb{C}$ is a linear map such that $\rho(a+ib)=\rho(a)+i\rho(b)$ for all self-adjoint elements $a,b\in B(H)$, then $\rho$ is a state precisely when its restrictions to the commutative subalgebras of $B(H)$ are states.

Both theorems show that the behaviour of maps on a C${}^*$-algebra is (to some extend) encoded by their localizations at the commutative subalgebras. This is one of the facts that are suggestive for Bohrification: whenever one is interested in maps that are local in some sense, classes of such maps can be best considered as presheaves on the corresponding local domain. The above theorems suggest that one is interested in presheaves on the poset of commutative subalgebras.

Indeed, Spitters et. al. \cite{hls09} assign to every C${}^*$-algebra $A$ (which we always assume to be unital) the poset $C(A)$ of commutative subalgebras of $A$, ordered by inclusion, and consider the topos $[C(A),\mathbf{Set}]$ of copresheaves on $C(A)$. This topos can now be endowed with an internal ring $\underline{A}$ given by the copresheaf 
\begin{diagram}
C(A) &  \rTo^{\; \; \; \underline{A}\; \; \;}  & \mathbf{Set} \\
C &  \rMapsto  & C
\end{diagram}
\noindent In fact, one can show that this internal ring is even an internal commutative C${}^*$-algebra.

This identification of a quantum phase space with a ringed topos allows the authors of \cite{hls09} to give a clear topos-theoretic picture of quantum kinematics. Using a constructive version of Gelfand duality \cite{cs082}, the internal ring (or commutative C${}^*$-algebra) $\underline{A}$ can be realized as the ring $\mathcal{C}(\underline{\Sigma}_{\: A},\mathbb{C})$ of continuous functions on an internal locale $\underline{\Sigma}_{\: A}$. An observable can then internally be described as a real-valued function on $\underline{\Sigma}_A$ and a state can be given by a probability density on $\underline{\Sigma}_{\: A}$. This internal description of quantum kinematics is quite similar to the description of classical kinematics, where an observable is given by a real-valued function $M\rightarrow \mathbb{R}$ on a manifold $M$ and states are usually described as probability densities on $M$. 

However, there is a significant difference between classical kinematics and this internal quantum kinematics. Classical kinematics has \emph{pure states}, whose probability densities are concentrated at one point of $M$. As shown in \cite{hlsw10}, it is a direct consequence of the Kochen-Specker theorem that the locale $\underline{\Sigma}_A$ has no global points, which means that these pure states do not arise in the internal description of quantum kinematics.\\

\noindent It is important to remark that the construction of a ringed topos from a C${}^*$-algebra extends to a \emph{contravariant} functor from a suitable category of C${}^*$-algebras to the category $\mathbf{RingTopos}$ of ringed toposes. To be more precise, we can only construct such a functor if we consider ${}^*$-homomorphisms between C${}^*$-algebras that reflect commutativity. A ${}^*$-homomorphism $h: A\rightarrow B$ is said to reflect commutativity if $h(a)$ and $h(b)$ commute in $B$, precisely when $a$ and $b$ commute in $A$.

Only these particular ${}^*$-homomorphisms can be used to construct morphisms of ringed toposes. Recall that a morphism $(\mathcal{E},\mathcal{O}_\mathcal{E})\rightarrow (\mathcal{F},\mathcal{O}_\mathcal{F})$ between ringed toposes consists of a geometric morphism $(f_*\vdash f^*): \mathcal{E}\rightarrow \mathcal{F}$ and a ring homomorphism $f^*\mathcal{O}_\mathcal{F}\rightarrow \mathcal{O}_\mathcal{E}$ in the topos $\mathcal{E}$. It is precisely the direction of this internal ring homomorphism that forces us to restrict our attention to the morphisms that reflect commutativity.

Indeed, a commutativity reflecting homomorphism $h: A\rightarrow B$ induces a functor
\begin{diagram}
C(B) & \rTo^{C(h)\;} & C(A)\\
D & \rMapsto & h^{-1}(D)
\end{diagram}
which in turn induces an essential geometric morphism $($Ran$_{C(h)}\vdash C(h)^*): [C(B),\mathbf{Set}]\rightarrow [C(A),\mathbf{Set}]$. Moreover, there is a canonical ring homomorphism
\begin{diagram}
C(h)^*\underline{A} & \rTo^{\epsilon} & \underline{B}\\
h^{-1}(D) & \rTo^{\epsilon_D = h\; \;} & D.
\end{diagram}
Here we see concretely why we need our morphisms to reflect commutativity and why we extend the Bohrification construction to a contravariant functor: if we chose our Bohrification functor to be covariant, we would essentially have to give all the ring homomorphisms in the opposite direction, which would only be possible if our ${}^*$-homomorphism were an embedding.

If we let $\mathbf{CStar}_{cr}$ be the category of C${}^*$-algebras with commutativity reflecting homomorphisms between them, we can therefore describe Bohrification as a functor $B: \mathbf{CStar}_{cr}^{op}\rightarrow\mathbf{RingTopos}$. We can refine this by noting that the topos $[C(A),\mathbf{Set}]$ is equivalent to the category of sheaves on a topological space: if $\mathcal{C}(A)$ is the poset $C(A)$ with the upwards closed subsets as its opens, then there is an equivalence between the categories $[C(A),\mathbf{Set}]$ and $\mathbf{Sh}(\mathcal{C}(A))$. We can therefore also say that Bohrification assigns to each C${}^*$-algebra $A$ a ringed space $(\mathcal{C}(A),\underline{A})$ and to each commutativity reflecting ${}^*$-homomorphism a map of ringed spaces, so that it is actually a functor
$$
B: \mathbf{CStar}_{cr}^{op}\rightarrow\mathbf{RingSp}\rightarrow\mathbf{RingTopos}
$$
to the category of ringed spaces.

The category of ringed spaces is easier to handle than the category of ringed toposes, especially in the computation of limits: a limit of ringed spaces consists of the limit of the underlying topological spaces, endowed with a colimiting sheaf of rings. However, we keep implicitly identifying each ringed space with the corresponding ringed topos.

\section*{Local nets}
\addcontentsline{toc}{section}{Local nets}
We will study the generalization of the previous constructions from quantum
mechanics to quantum field theory and from plain quantum kinematics to
quantum dynamics, using algebraic quantum field theory (AQFT) to describe a QFT.
The idea of AQFT is to characterize a quantum field theory by the assignment of algebras of local observables to each open subset of $X$. These observables present what can be measured by performing an experiment within that certain region of space and time. 

The spacetime $X$ is given by a Lorentzian manifold, so that the tangent space in each point $x$ decomposes in a spacelike and a timelike region, separated by the lightcone of vectors whose length is $0$. We say that two points are \emph{spacelike separated} if there is no timelike or lightlike curve between them. In Minkowski space $\mathbb{R}^{1+n}$, this means that the straight line between the two points is spacelike. Two subsets $U$ and $V$ of $X$ are spacelike separated if all $x\in U$, $y\in V$ are spacelike separated.

In this way, each open $O$ in $X$ has a causal complement $O'$ consisting of points that are spacelike separated from $O$. We say that $O$ is causally complete if $O''=O$ and denote by $\mathcal{V}(X)$ the poset of causally complete opens, ordered by inclusion. In two-dimensional Minkowski space, $\mathcal{V}(X)$ consists of the causal diamonds and the left and right wedges. 

AQFT now describes a quantum field theory as a copresheaf $A: \mathcal{V}(X)\rightarrow \mathbf{CStar}_{inc}$ to the category of C${}^*$-algebras with inclusions between them. This reflects the idea that within more space and time, more observables can be measured. Moreover, one imposes the condition that for any two spacelike separated opens $O_1$ and $O_2$ in $X$, the algebras $A(O_1)$ and $A(O_2)$ mutually commute in $A(O_1\vee O_2)$, where $O_1\vee O_2$ denotes the smallest causally complete open containing both $O_1$ and $O_2$. This condition basically says that relativistic independence (i.e. regions being spacelike separated) should imply quantum mechanical independence (i.e. that two observables from separated regions commute). A copresheaf $A: \mathcal{V}(X)\rightarrow \mathbf{CStar}_{inc}$ that satisfies this causal locality condition will be called a local net.

In Minkowski space, the causal locality of a net has an important equivalent formulation in terms of its restriction to a Cauchy surface. A connected hypersurface $S$ in a Lorentzian space is said to be a \emph{Cauchy surface} if every timelike or lightlike curve intersects $S$ in precisely one point. This condition formalizes the idea that the points on $S$ give a space at one specific time. In Minkowski space, any plane spanned by only spacelike vectors forms a Cauchy surface.

One can restrict a local net $A$ to a Cauchy surface $S$ by noting that each connected open $U\subseteq S$ is contained in a smallest causally complete open $O_U\subseteq X$. The restriction of a net $A$ to $S$ will then be the net $A|_S: \mathcal{V}(S)\rightarrow\mathbf{CStar}_{inc}$ that sends each $U$ to $A(O_U)$. With these definitions, one easily verifies that for Minkowsi space $X$, a net $A: \mathcal{V}(X)\rightarrow \mathbf{CStar}_{inc}$ is local precisely when its restriction $A|_S$ to any Cauchy surface $S$ has the property that for two disjoint opens $U, V\subseteq S$ their algebras $A|_S(U)$ and $A|_S(V)$ mutually commute. We use this alternative formulation of causal locality when we discuss the characterization of locality in terms of Bohrified nets.

\section*{Bohrified nets}
\addcontentsline{toc}{section}{Bohrified nets}
The formalism of AQFT thus describes a quantum field theory by a local net $A: \mathcal{V}(X)\rightarrow \mathbf{CStar}_{inc}$. Since inclusions of C${}^*$-algebras certainly reflect commutativity, we can compose this local net with the Bohrification functor $\mathbf{CStar}_{cr}^{op}\rightarrow \mathbf{RingSp}$. This composition gives a presheaf $B(A): \mathcal{V}(X)^{op}\rightarrow \mathbf{RingSp}$, which we will call the Bohrified net. 
It sends an open $O$ to the topos $[C(A(O)),\mathbf{Set}]$, endowed with its tautological internal ring. For an inclusion of opens $O_1\subseteq O_2$,  the local net $A$ gives an inclusion of C${}^*$-algebras $A(O_1)\subseteq A(O_2)$. The corresponding morphism of ringed spaces $B(A)(O_2)\rightarrow B(A)(O_1)$ then consists of a geometric morphism
$$
[C(A(O_2)),\mathbf{Set}] \pile{\lTo \\ \bot \\\rTo} [C(A(O_1)),\mathbf{Set}]
$$
and a morphism of internal rings. Following the construction of the Bohrification functor, we see that the geometric morphism is induced by the functor $C(A(O_2))\rightarrow C(A(O_1))$ that sends a commutative subalgebra $C\subseteq A(O_2)$ to the intersection $C\cap A(O_1)$.

If $O$ is an open of $X$, then the Bohrified net contains all information about whether elements commute in $A(O)$. This suggests that the causal locality of the original local net $A$ should be reflected in terms of its Bohrified net $B(A)$. Indeed, we will show that the causal locality of a net of observables $A$ is related to a descent condition on its Bohrified net $B(A)$. To do this, we will impose two conditions on our net of observables that often arise in discussions of AQFT (see for instance \cite{ha96}): we will require our nets of observables to be both {additive} and {strongly local}. 

A net $A$ is said to be \emph{additive} if for any two spacelike separated opens $O_1$ and $O_2$ such that $\overline{O_1}\cap\overline{O_2}\neq \emptyset$, one has that $A(O_1\vee O_2)=A(O_1)\vee A(O_2)$; the algebra $A(O_1\vee O_2)$ is generated by $A(O_1)$ and $A(O_2)$ in $A(X)$. Being additive in some sense expresses the local character of AQFT: the observables in two small opens $O_1$ and $O_2$ suffice to describe everything that can be observed in the larger open $O_1\vee O_2$ generated by them. For nets on $\mathbb{R}^2$, this condition follows from the so-called \emph{split-property for wedges}, as discussed in \cite{gl06}. 

A net $A:\mathcal{V}(X)\rightarrow\mathbf{CStar}_{inc}$ is \emph{strongly local} if it is causally local and has the property that for any two spacelike separated opens $O_1$ and $O_2$, and any pair of commutative subalgebras $C_1\subseteq A(O_1)$ and $C_2\subseteq A(O_2)$, one finds for the algebra $C_1\vee C_2\subseteq A(O_1\vee O_2)$ generated by them that $(C_1\vee C_2)\cap A(O_1)=C_1$ and $(C_1\vee C_2)\cap A_2=C_2$.

Strong locality is precisely the kind of locality condition we need in our main theorem. This condition holds for nets satisfying \emph{Einstein causality}, which is one of the axioms for an AQFT imposed in \cite{bf09}. 
A net $A:\mathcal{V}(X)\rightarrow\mathbf{CStar}_{inc}$ is called Einstein causal if for any two spacelike separated opens $O_1$ and $O_2$, one has that the inclusions $A(O_1)\subseteq A(O_1\vee O_2)$ and $A(O_2)\subseteq A(O_1\vee O_2)$ factor over the tensor product
\begin{diagram}
 & & A(O_1\vee O_2) & &\\
& \ruInto & \uInto & \luInto & \\ 
A(O_1) & \rInto & A(O_1)\otimes A(O_2) & \lInto & A(O_2).
\end{diagram}

In \cite{bf09} it is argued that Einstein causality expresses that the subsystems localized at $O_1$ and $O_2$ are completely independent: ordinary locality only states that it does not matter whether you first do a measurement in $O_1$ and then one in $O_2$, or the other way around. Einstein causality adds to this that the subsystems localized at $O_1$ and $O_2$ are even statistically independent: a state $\rho_1$ on the system at $O_1$ and a state $\rho_2$ on the system at $O_2$ give a product state $\rho_1\otimes\rho_2$ on composite system.

\noindent Our main result relates the above two conditions on a local net $A$ to a more geometric condition on its Bohrified net $B(A)$. In particular, we find a natural way to express the causal locality of a net $A$ in terms of a descent condition on $B(A)$.
\begin{ntheorem}
An additive net $A: \mathcal{V}(\mathbb{R}^{1+n})\rightarrow\mathbf{CStar}_{inc}$ on a Minkowski spacetime is strongly local, precisely when the restriction of its Bohrified net $B(A)$ to any Cauchy surface satifies descent by \emph{local} geometric surjections. 
\end{ntheorem}

\noindent Recall that a local geometric morphism $(f_*\vdash f^*): \mathcal{E}\rightarrow\mathcal{F}$ is a geometric morphism such that the direct image $f_*$ has a further right adjoint which is full and faithful (cf. \cite{joh02} section C3.6). It follows directly from its definition that a local geometric morphism is always a geometric surjection. Local geometric morphisms model `infinitesimal thickenings': if a sheaf topos $\mathbf{Sh}(X)$ has a local geometric surjection to the point $\mathbf{Sh}(*)=\mathbf{Set}$, then $X$ is the `infinitesimal thickening' of a point, in the sense that there is a point whose only neighbourhood is the whole space $X$. A strongly local net thus induces a Bohrified net which is `infinitesimally' close to being a sheaf.

We remark that this result clarifies one particular curiosity in the formalism of AQFT: it is a strange aspect of AQFT that a local net does not satisfy a descent condition of any kind. Indeed, whenever one considers a (co)presheaf on a topological space, one usually demands it to be a sheaf. However, we have seen that the basic axioms for a local net on a spacetime $X$ do not make a single use of the covering relation on the opens of $X$. 

By adding the additivity condition to the axioms of a local net, we did use coverings in the axioms of AQFT. Still, even the additivity condition does not translate directly into a descent condition on the local net $A: \mathcal{V}(X)\rightarrow\mathbf{CStar}_{inc}$ (see for instance \cite{fh87}). In fact, it is typically hard to combine the causal locality of the net $A$ with any kind of descent of $A$. Our result now shows that one can one can naturally impose a descent condition on the Bohrified net $B(A)$, instead of the original net $A$. In fact, we see that the (strong) locality of the net $A$ is actually necessary to establish the descent of $B(A)$ by local geometric morphisms. The statement therefore exhibits a tighter link between AQFT and the geometry of the spacetime $X$ than becomes clear from looking at local nets only.

For an idea of the proof, let $A$ be an additive, strongly local net and let $S\subset \mathbb{R}^{1+n}$ be some Cauchy surface. We consider the case where just two connected opens $U$ and $V$ cover their union $U\cup V$ in $S$. We then find a descent morphism
$$
B(A)|_S(U\cup V) \rightarrow B(A)|_S(U) \times_{B(A)|_S(U\cap V)} B(A)|_S(V)
$$
in the category of ringed spaces. Forgetting the ring structure, this descent morphism consists of a geometric morphism $f^*\dashv f_*$ between the corresponding toposes of copresheaves on the commutative subalgebras. 

Because the codomain of the Bohrified net $B(A)$ is the category of ringed spaces, we find that this geometric morphism $f^*\dashv f_*$ is ultimately induced by a functor $f$ on the posets of commutative subalgebras. In this case, the functor $f$ is given by the functor
\begin{diagram}
C(A|_S(U\cup V)) & \rTo^{\;f\;\;} & C(A|_S(U)) \times_{C(A|_S(U\cap V))} C(A|_S(V))
\end{diagram}
that sends a commutative subalgebra $C\subseteq A|_S(U\cup V)$ to the pair of intersections $(C\cap A|_S(U), C\cap A|_S(V))$. 

If $U$ and $V$ are disjoint, then the locality of $A$ implies that $A|_S(U)$ and $A|_S(V)$ mutually commute in $A|_S(U\cup V)$. This means that for any pair of commutative subalgebras $C_1\subseteq A|_S(U)$, $C_2\subseteq A|_S(V)$, there is a smallest commutative subalgebra $C_1\vee C_2$ in $A|_S(U\cup V)$ containing both $C_1$ and $C_2$. This construction gives a functor 
\begin{diagram}
C(A|_S(U)) \times_{C(A|_S(U\cap V))} C(A|_S(V)) & \rTo & C(A|_S(U\cup V))
\end{diagram}
which is easily checked to be a left adjoint to $f$. In fact, if $A$ is a strongly local net, then this left adjoint is even full and faithful.

We can then lift this adjunction to the level of geometric morphisms: the fact that $f$ has a left adjoint implies that the descent morphism $f^*\dashv f_*$ has a further right adjoint. If the left adjoint to $f$ is full and faithful, then this extra right adjoint to $f_*$ is actually full and faithful, which means precisely that the descent morphism $f^*\dashv f_*$ is a local geometric morphism. We thus find that the Bohrified net satisfies descent by a local geometric morhism if we have a cover of disjoint opens.

In the case that $U$ and $V$ are not disjoint, the additivity of the local net allows us to use $U\cap V$ and the interiors of $U\setminus V$ and $V\setminus U$ instead of $U$ and $V$ to `cover' the union $U\cap V$. Indeed, the additivity condition of the local net $A$ states that $A|_S(U)$  is generated by $A|_S(U\cap V)$ and $A|_S(U\setminus V^{o})$. We can therefore essentially replace $U$ and $V$ by these three disjoint opens, and apply the construction for the case in which $U$ and $V$ were disjoint. The result is again that $f$ has an extra left adjoint, which is full and faithful, so that the geometric morphism $f^*\dashv f_*$ is local.

Summarizing, we find for an additive, strongly local net $A$ that the restriction of the Bohrified net $B(A)$ to a Cauchy surface satifies descent by local geometric surjections. Conversely, if the restriction $B(A)|_S$ satifies this descent condition, then one can deduce that $A|_S(U)$ and $A|_S(V)$ mutually commute if $U$ and $V$ are disjoint opens of the Cauchy surface $S$. From this it follows that the net $A$ is local and actually even strongly local.

\section*{Conclusion}
\addcontentsline{toc}{section}{Conclusion}
We have related the causal locality of a net of observables to a descent condition on the corresponding Bohrified net.
This interplay between the causal locality of nets of observables on one hand, and on the other hand the locality of the Bohrified net in the sense of sheaf theory, gives a relation between the topology of the spacetime and the theory of AQFT. By assigning an active role to the spacetime geometry, it gives AQFT a much more geometric flavour, which will be particularly important if one tries to consider local nets on curved spacetimes, as is done in \cite{bf09}.

Our result might therefore add to the insights by D\"oring-Isham and Spitters et. al. that many aspects of quantum theory have a natural formulation when one models a quantum phase space as a ringed topos. In particular, it suggests that the ideas by these two groups might have some useful applications in quantum field theory.

\section*{Acknowledgements}
\addcontentsline{toc}{section}{Acknowledgements}
I am grateful to Urs Schreiber for taking the time to comment on a draft of this text. I thank the referees for their comments and suggestions.

\end{document}